\documentclass[preprint,showpacs,preprintnumbers,amsmath,amssymb]{revtex4}

\usepackage{graphicx,psfrag}
\usepackage{dcolumn}
\usepackage{bm}

\begin{document}

\preprint{PUPT-2295}

\title{Nanogap Transducer for Broadband Gravitational
Wave Detection}

\author{Guilherme L. Pimentel}
 \altaffiliation[Also at ]{Departamento de F\'isica, Instituto Tecnol\'ogico de
 Aeron\'autica.}
 \email{gpimente@princeton.edu}
\affiliation{%
Joseph Henry Laboratories, Princeton University, Princeton, NJ, 08544.
}%

\author{Odylio D. Aguiar}
 \email{odylio@das.inpe.br}
 \homepage{http://www.das.inpe.br/graviton/index.html}
\affiliation{
Divis\~ao de Astrof\'isica, Instituto Nacional de Pesquisas Espaciais, Jd. da Granja,
 S\~ao Jos\'e dos Campos, SP, Brazil, 12201-970.
}%

\author{Michael E. Tobar}
\email{mike@physics.uwa.edu}
\affiliation{School of Physics, The University of Western Australia, Crawley, WA, Australia,  6009.}

\author{Joaquim J. Barroso}
 \email{barroso@plasma.inpe.br}
\affiliation{
Laborat\'orio Associado de Plasma, Instituto Nacional de Pesquisas Espaciais, Jd. da Granja,
 S\~ao Jos\'e dos Campos, SP, Brazil, 12201-970.
}%

\author{Rubens M. Marinho Jr.}
 \email{marinho@ita.br}
 \affiliation{Departamento de F\'isica, Instituto Tecnol\'ogico de Aeron\'autica, Campo Montenegro, S\~ao Jos\'e dos Campos, SP, Brazil, 12228-000.}

\date{\today}

\begin{abstract}
By changing from a resonant multimode paradigm to a free mass paradigm for transducers in resonant mass gravitational wave detection, an array of six spheres can achieve a sensitivity response curve competitive with interferometers, being as sensitive as GEO600 and TAMA300 in the $3-6$ kHz band and more sensitive than LIGO for 50\% of the $6-10$ kHz band. We study how to assemble a klystron resonant cavity that has a 1 nm gap by understanding the stability of the forces applied at it (Casimir force, elastic force, weight). This approach has additional benefits.  First, due to the relatively inexpensive nature of this technology ($\sim$US\$ 1 million), it is accessible to a broader part of the world's scientific community.  Additionally, spherical resonant mass detectors have the ability to discern both the direction and polarization resolutions. 
\end{abstract}

\pacs{Valid PACS appear here}
\maketitle

\section{\label{sec:level1}Introduction}

In the field of experimental gravitation, the detection of gravitational 
waves (GW) is a major challenge. Indirect evidence of GW emission
was discovered by Hulse and Taylor \cite{HULSE:1975p7227}, but no direct detection has yet been made.
Searching for GW is important in terms of probing new physics, as well as testing Einstein's General Relativity. In terms of sources detectable with ground-based detectors, see references \cite{Kokkotas:1999p6829} and \cite{Cutler:2002p6830}. Resonant-mass (RM) detection, pioneered by Weber \cite{Weber:1960p6818}, has the major drawback of small bandwidth.

\begin{figure}[h]
\begin{center}
\includegraphics[width=.4\textwidth]{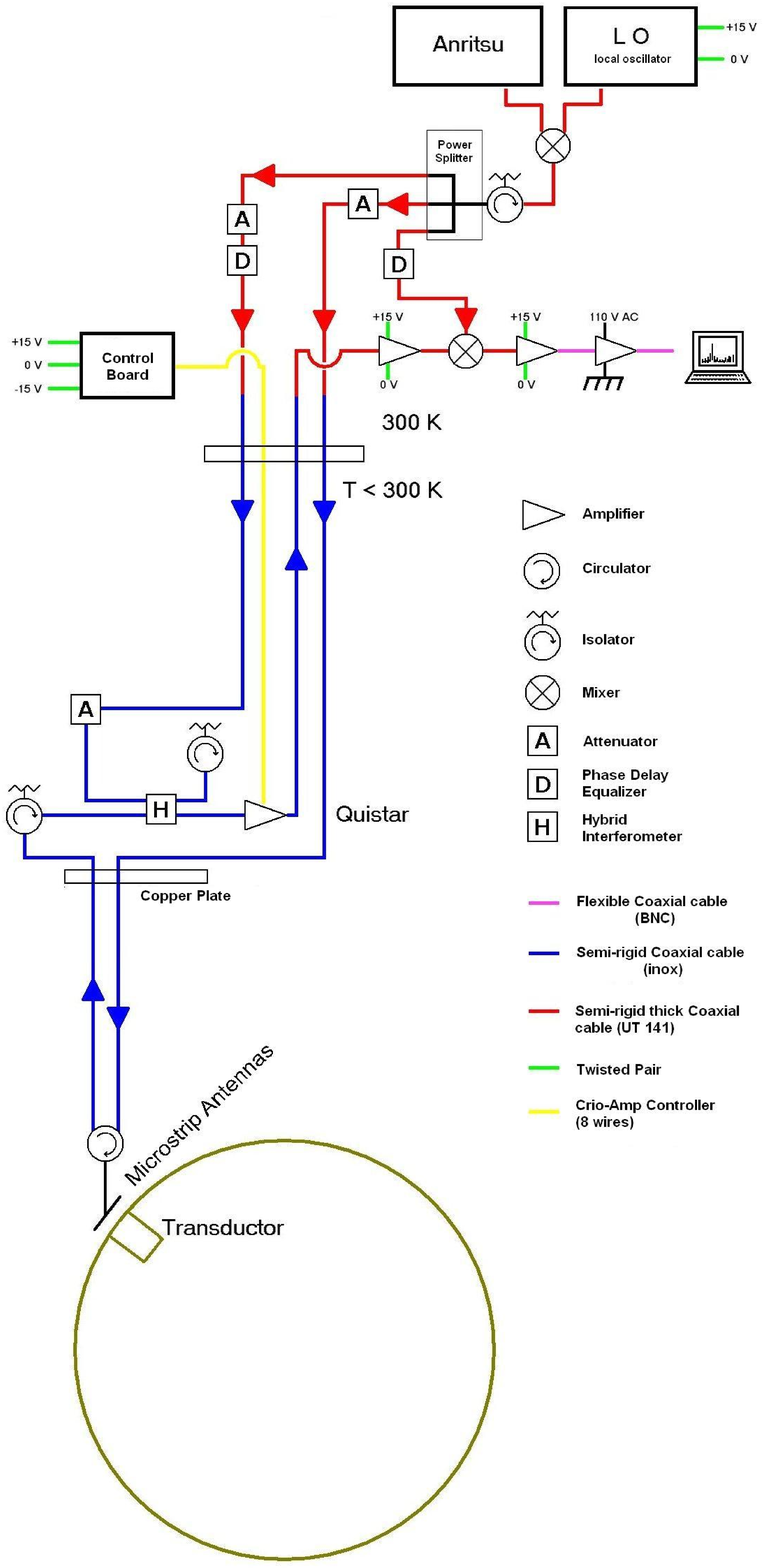}
\caption{(Color online). Block diagram for Schenberg.}
\label{blockdiagram}
\end{center}
\end{figure}

We address this question in the context of the Mario Schenberg GW detector \cite{Aguiar:2008p313}. Figure \ref{blockdiagram} shows a schematic with all the electrical wiring and block diagram for each transducer of Schenberg. These transducers are arranged according the truncated icosahedron configuration proposed in \cite{Johnson:1993p9003}.

We propose a novel approach to the design of resonant 
cavities for RM detectors. We devise such a cavity for Schenberg, proposing a way to build 
it, as well as discussing technological issues. The main idea is reducing a gap between two surfaces of the cavity down to $1$ nm.

We also discuss how to achieve the desired gap, by studying the forces acting on the cavity's membrane. They are the elastic force, the Casimir force, and the membrane weight. We see that for our specific case the Casimir force may scale as $\sim g^{-3}$, instead of $\sim g^{-4}$, where $g$ is the gap in the cavity under discussion. 

The paper is organized as follows: in section II, we briefly review the formalism of \cite{Bonaldi:2006p6716} and apply it to an array of six spheres that could be competitive compared to the interferometric detectors in the $3-10$ kHz frequency range. In section III we address the question on how to build the cavity, and in section IV our concluding remarks.

\section{Sensitivity Curve of the Schenberg Detector}

We now calculate the sensitivity 
curve of the Schenberg detector. We follow the approach presented in \cite{Bonaldi:2006p6716}.

The sensitivity curve of a gravitational wave detector is defined as the strain (or amplitude)
noise  spectral density (which is the square root of the power spectral density) 
when there is no signal present in the detector, i.e., the 
response of the detector to different noise sources that will act on it. We model our spherical detector using the following noise sources:

\begin{itemize}
\item Series noise: Noise from the chain of ampliЮers in series, which is dominated by 
the pre-amplifier; 
\item Phase noise: Noise from the sidebands of the microwave pump signal; 
\item Seismic noise: Noise from seismic movements of the detector's site; 
\item Backaction noise: Noise from the amplitude noise of the microwave pump, which 
causes a backaction force on the transducer. 
\item Thermal noise: Noise from the Brownian motion of the sphere's atoms.
\end{itemize}

These noise sources are either of 
``displacement"~type, i.e., they produce small 
displacements on the surface on which they act (series noise, phase noise, thermal and seismic noise), or of ``force"~type, such as backaction noise.

So, in general, there will be $S_{ff}(\omega)$ and $S_{xx}(\omega)$, force 
and displacement noise spectral densities. We want to obtain $\sqrt{S_{hh}(\omega)}\equiv\tilde{h}(\omega)$ from these, 
i.e., the strain noise sensitivity curve.
We write our strain noise sensitivity as:
\begin{equation}
\tilde{h}(\omega) = \sqrt{ \frac{S_{xx}(\omega) + S_{ff}(\omega) \mid H_{BA}(
\omega)\mid ^2}{\mid H_{GW}(\omega)\mid ^2}}
\end{equation}
where $H_{BA}(\omega)$ is the transfer function between 
the displacement the system (sphere) suffers and a force applied at the 
transducer's position, i.e. $H_{BA}(\omega)=\frac{\tilde{X}(\omega)}{\tilde{F}(\omega)}$
and $H_{GW}(\omega)$ is the transfer function between the displacement of the 
sphere at the transducer's position and the amplitude of the metric 
perturbation, i.e. $H_{GW}(\omega)= \frac{\tilde{X}(\omega)}{\tilde{h}(\omega)}$.

When driven by a force, a resonant detector responds with small oscillations 
throughout its volume. The displacement vector $\mathbf{u}(\mathbf{r},t)$ 
obeys the following equation:
\begin{equation}
\rho \frac{\partial ^2 \mathbf{u}(\mathbf{r},t)}{\partial t^2} - \mu \nabla ^2 
\mathbf{u}(\mathbf{r},t)-(\lambda + \mu)\nabla (\nabla \cdot \mathbf{u}(
\mathbf{r},t)) = \mathbf{f}(\mathbf{r},t)
\label{eigenvalueproblem}
\end{equation}
where $\lambda$ and $\mu$ are the Lam\`e's coefficients, $\rho$ is the sphere's 
density, and $\mathbf{f}(\mathbf{r},t)$ is the force density applied to the 
sphere. We will be interested in factorizable forms of the force density, 
i.e., those which can be written as a product of a spatial term and a time 
term. In this case the treatment in terms of eigenmodes of the sphere is 
straightforward, and is done in detail in \cite{Lobo:1996p6430,Lobo:2000p6417}. 

Now, we move on to the construction of the gravitational wave transfer 
function. We must be able to express the output variable as well as the force 
input. We begin with the latter.

For an incoming gravitational wave along the z-axis, the time and spatial part 
of the force density will be given by $
\mathbf{f_r}(r,\theta ,\phi)=(r \sin^2\theta \cos(2 \phi + \psi ), 
r \sin\theta \cos\theta \cos(2 \phi +\psi), -r \sin\theta \sin (2 \phi +\psi))$ and $
f_t(t)=\frac{1}{2} \rho \ddot{h}(t)$, where $\phi$, $\theta$ and $\psi$ are the azimuthal and polar spherical coordinate angles and GW polarization angle, respectively.
As each mode of the sphere is treated as a harmonic oscillator, with resonant 
frequency given by the solution of the eigenvalue problem stated in equation 
\ref{eigenvalueproblem} and constant Q-factor (by assumption), equal 
to the sphere's ($1.4 \times 10^6$), the transfer function will have a sum 
over each mode with a ``harmonic oscillator" -like term.

Now we move on to the specification of the observable variable. In the case 
of the sphere, we will observe the displacement at its surface, and measure 
the radial displacement at a given point, specified by the pair 
$(\Theta,\Phi)$. After we find the sensitivity curve for a specified position, 
we will average the sensitivities over all possible solid angles. If the GW 
is always coming from the z-axis, this solid angle average 
is equivalent to maintaining the detector's position fixed and changing the 
position of incidence of the GW. It is also important to note that from a 
symmetry argument this analysis also is polarization independent, as a 
suitable rotation through the z-axis takes a particular polarization to 
either the $\times$ or $+$ polarization. (We are just interested in the 
quadrupole modes, as the monopole and dipole modes are not excited by GR, and 
the spherical detector will only couple to the quadrupole modes, and none of 
the higher-order modes \cite{Gasparini:2005p4661})

The gravitational wave transfer function is given by (as shown in \cite{Bonaldi:2006p6716}):
$
H_{GW} (\omega) = \frac{1}{2V} \displaystyle\sum_n \frac{- \omega^2 \left[\int_V dV 
\mathbf{f_r}(\mathbf{r}) \cdot \mathbf{u_n}(\mathbf{r})\right ] \left[\int_S 
ds \mathbf{P}(\mathbf{r}) \cdot \mathbf{u_n}(\mathbf{r})\right ]}{(\omega_n^2 - 
\omega^2) + \imath \omega_n^2 /Q}
$. For the backaction transfer function, we write the backaction force 
acting on the transducer as proportional to the ``weight function", with the 
full force separable in a time and spatial part, just as we have done with the 
gravitational wave force previously. Making 
$\mathbf{F_{BA}}(\mathbf{r},t)=F_{BA}(t)\mathbf{P}(\mathbf{r})$, and after some algebra, we find the following: the backaction transfer function is given by $H_{BA}(\omega) = \frac{1}{M} \displaystyle\sum_n \frac{\left[ \int_S ds \mathbf{P}(\mathbf{r})
\cdot \mathbf{u_n}(\mathbf{r}) \right ] ^2}{(\omega_n^2 - \omega^2) + \imath \omega_n^2 /Q}$, where $\mathbf{P}(\mathbf{r})=\left(\frac{1}{R^2} \delta (\Omega - (\Theta , \Phi )),0,0 \right)$ has to do with the fact that we observe the readout of the detector at certain positions $\Theta$ and $\Phi$ (those are not physically relevant for the joint sensitivity because of the symmetry of a spherical detector). $\omega_n$ is the resonant frequency of the $nlm$ mode of the sphere. The summation runs over $n$, $l$ and $m$ and is implicitly stated.

In order to make a suitable model for the Brownian Noise, we use the 
fluctuation-dissipation theorem, which states that (at the classical level, the 
quantum corrections for our case are negligible):
$S_{brownian}(\omega) = -\frac{4 k_B T}{\omega}\Im m \left[ H_{BA} (\omega) 
\right] \nonumber = -\frac{4 k_B T}{\omega} \displaystyle\sum_n \frac {-\omega_n^2}{M Q}
\frac{\left| \int_S ds \mathbf{P}(\mathbf{r})
\cdot \mathbf{u_n}(\mathbf{r}) \right | ^2}{(\omega_n^2 - 
\omega^2)^2 + \omega_n^4 /Q^2}$
For the various noise sources mentioned before, we use expressions previously derived elsewhere in the literature \cite{Tobar:2000p7206}, together with the thermal Brownian noise expression just shown:
\begin{eqnarray}
S_{series}(\omega)&=& \displaystyle L_{amp} \frac{k_B \left(T_{amp}+T\right)}{P_{inc}}\left( \frac{2 Q_e}{f_0}
\frac{df_0}{dg} \right)^{-2} \label{noisefirst}\\
S_{phase}(\omega)&=& \displaystyle S_{ph} \left(\frac{df_0}{dg} \frac{2 \pi}{\omega}\right)^{-2}\\
S_{seismic}(\omega)&=&\displaystyle\frac{a (2 \pi)^4}{\omega^4}\label{seismicn}\\ \nonumber \\
S_{backaction}(\omega)&=& \displaystyle\frac{P_{inc}^2}{2 \omega_0 ^2} \left( \frac{2 Q_e}{f_0}
\frac{df_0}{dg} \right)^2 S_{am}
\end{eqnarray}
Where: $L_{amp}$ represents the losses in the transmission system from the 
transducer cavity to the microwave cryogenic amplifier (the one inside the dewar); $k_B$ is Boltzmann's constant; $T_{amp}$ is the amplifier's noise temperature, which is currently $\sim 10$ K but, in principle, can be lowered to $\sim 1$ K
\cite{HEFFNER:1962p7173}, $T$ is the thermodynamic temperature of the system, which 
is currently liquid helium (about $4$ K) but can be lowered 
down to $\sim 10$ mK with use of dilution refrigeration; $P_{inc}$ is the power which the microwave pump delivers at the cavity's input, which can be calculated for a specific cavity geometry, and in our case will be of $\sim 10^{-10}$ W; $Q_e$ is the electrical Q-factor of the cavity, which is estimated to be $1.5\times 10^6$ when the cavity is cooled down to $\sim 10$ mK, where 
Niobium is a better superconductor than at $4$ K ($T_c=9.3$ K \cite{Karakozov:2004p1814}); $f_0$ is the microwave pump frequency, $\simeq 10$ GHz; $\displaystyle\frac{df_0}{dg}$ is the variation of the cavity's resonant frequency with small gap displacements, and our main goal is to achieve for this 
parameter a value of $500$ MHz/\AA; $S_{ph}$ is the phase noise for our amplifier, which is input power 
dependent. Values of $-190$ dBc/Hz are technologically available currently \cite{Tobar:2000p7206}; $a$ is a site-dependent parameter. Typical values for ground detectors are 
$10^{-16}$ m$^2$/Hz; $M$ is the sphere's mass; $Q$ is the mechanical Q-factor of the sphere.

\begin{figure}[h]
\begin{center}
\includegraphics[width=90mm]{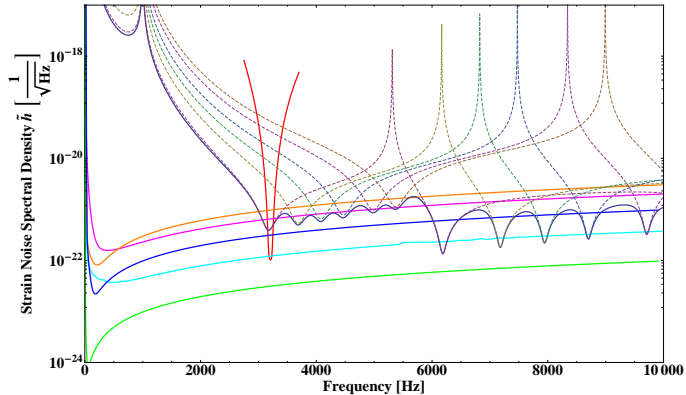}
\caption{(Color online). The sensitivity curve for the Schenberg detector depicting its components. The dashed curves represent each of the 6 spheres we chose for the array (masses: 1150 kg (Schenberg), 744 kg, 547 kg, 414 kg, 301 kg, 239 kg), the lowest frequency being Schenberg. The V-shaped red curve is Schenberg with its usual configuration operating at dilution fridge temperatures ($\sim10$ mK). Interferometer curves are also plotted: advanced LIGO (green), LIGO (blue), VIRGO (light blue), TAMA300 (pink) and GEO600 (orange). All of these are project curves, not actual data.}
\label{sensitivityschenberg}
\end{center}
\end{figure}

Now we should analyze those equations in the specific case of Schenberg, which 
will yield the sensitivity curve shown in Figure \ref{sensitivityschenberg}. 
From the Figure we can note that there is
good sensitivity near the quadrupole-mode resonances (which are the only 
modes which couple to the wave \cite{Lobo:2000p6417,Gasparini:2005p4661}).  
The series noise is the main contribution to the 
sensitivity curve, and the only way we could make this curve better would be 
by pumping more power into the cavity, that would lead it into electrical 
breakdown. It seems that this is the technological limit available. At 
low frequencies the seismic noise dominates, but the vibration isolation 
system we use is devised to have large attenuation at high frequencies (
$S_{seismic}(\omega)$ will be attenuated by a factor $(\omega/\omega_0)^{2N}$ \cite{MICHELSON:1987p7222}, 
where $\omega_0$ is designed to be $\sim 50$ Hz and $N=5$ because we have 
five mechanical filters in series).

In order to outcome the problem of low sensitivities between consecutive 
quadrupole modes of the detector, we propose a spherical 
antenna array that would be competitive with interferometric detectors 
in the range $3\sim10$ $kHz$.

The product $R\omega_{nlm}$ is a constant for a sphere when we deal with its 
resonant modes, in particular, its quadrupole modes \cite{Lobo:2000p6417}. If we 
think of an array of spheres, instead of only one sphere, we can combine the 
sensitivity curves of a sphere array and choose our spheres' radius so that 
one sphere's quadrupole modes hide the weakness of the band 
between another sphere's consecutive quadrupole modes, at which its sensitivity is poor. 
This has an advantage over an array of interferometers because the 
correlation coefficient is unity, i.e., a signal which excited one sphere 
must, in principle, excite all of the spheres from the array, because of 
their omnidirectionality, when in the case of interferometers we must take 
into account their relative orientations, which introduces intrinsic losses in their 
correlation functions.

To compose the sensitivities, note that as we look for simultaneous events in uncorrelated detectors, their occurrence must go as $e^{-E/kT_1}e^{-E/kT_2}...e^{-E/kT_n}\sim e^{-E/kT_{eff}}$
 so $\frac{1}{T_{eff}}=\sum_i\frac{1}{T_i}$. As $T_i\sim h_i^2$, $\frac{1}{\tilde{h}_{eff}(\omega)}=\sqrt{\sum_i\frac{1}{\tilde{h}^2_i(\omega)}}$. This is how we compose sensitivities for the result shown in Figure \ref{sensitivityschenberg}.
 
The plot is linear-log in order to highlight the band of frequencies in which 
we propose to be competitive (from $3$ to $10$ kHz). This array of spheres can be competitive against 
interferometers if we tune the resonant frequencies 
of the spheres side-by-side.

A few remarks:
\begin{itemize}
\item We suppose that, besides the sphere's mass and radius, i.e., the 
distribution of its resonant modes, all of the other parameters are kept 
unchanged (thermodynamic temperature, electronic system parameters, etc.);
\item The low frequency behaviour can be enhanced by a proper project for the 
vibrational isolation system. This was not the case previously because we 
were aiming only for the Schenberg detector's narrow band behaviour.
\item The parameters used for the simulation shown in figure \ref{sensitivityschenberg} 
were:
\begin{itemize}
\item $S_{am}=-195$ dBc/Hz; $S_{ph}=-185$ dBc/Hz;
\item $\displaystyle\frac{df_0}{dg}=5 \times 10^{18}$ Hz/m;
\item $Q=1.4\times10^6$; $Q_e=1.5\times10^6$;
\item $f_0=10$ GHz;
\item $T_{amp}=1$ K; $T=10$ mK;
\item $M_{spheres}$  (kg) = $1150$ (Schenberg), $744$, $547$, $414$, $301$, $239$.
\end{itemize}

$S_{am}$ and $S_{ph}$ were designed from a room temperature extrapolation, for 10 GHz \cite{Tobar:2000p7206}, and are now being built. Electrical quality factors of $10^9$ were reported for niobium superconductive cavities \cite{Geng:2005p15881, Proch:1998p15882}. In the case of klystron cavities the highest value measured was $\sim$500 k \cite{Locke:2001p15708}, but we simulated with the SuperFish code the possibility to reach values as high as $1.5 \times 10^6$, considering values of superconductivity at $10$ mK. The mechanical Q factor of $1.4 \times 10^6$ is conservative, because we have already measured for the sphere Qs as high as $2 \times 10^6$ at 2K \cite{Aguiar:2004p15627}. We degraded the value assuming some dissipation factor at the transducer's silicon/niobium membrane, which will be confirmed during the development of the membranes for the standard resonant-transducers. The carrier used is at 10 GHz, which is a common value for operation. About noise temperature, we have amplifiers from Quinstar with $T_{amp} \sim 8K$ at 4.2 K of thermodynamical temperature and 10 GHz of operating frequency, but it might be possible to reach a noise temperature of 1 K at 10 mK and 10 GHz. The American National Radio Astronomy Observatory \cite{Staggs2009} measured a noise temperature of 2 K at 12.5 K and 10 GHz. Finally, the masses 744, 547, 301, and 239 kg were chosen in order to optimize the sensitivity curve of the array (High sensitivity peaks where the other spheres have low sensitivity). These are, of course, smaller spheres than Schenberg, which are, therefore, feasible and less expensive to build.

\item We can improve the low frequency sensitivity introducing larger spheres 
in the array, as well as improving the seismic noise attenuation system. This 
is not proposed here because we aim at a viable, large collaboration between 
different groups, with different budgets, and enlarging the sphere would 
require larger costs.
\end{itemize}

\section{Designing the Resonant Cavity}

A klystron cavity is a special type of reentrant cavity. Its name is due to
the possibility of producing electric fields with large amplitudes in a small
volume of the cavity, which can be used to control electron beams, and is used
in klystron tubes. The choice for a geometry is based on the number of adjustable 
parameters, as well as the feasibility of the cavity. Based on Fujisawa's work \cite{Fujisawa:1958p5806}, 
which uses a semi-empirical method based on field configurations inside the cavity, 
as well as a lumped parameter equivalent circuit, we can design our cavity 
based on its conductivity and dimensions.

\begin{figure}[h]
\begin{center}
\includegraphics[width=.4\textwidth]{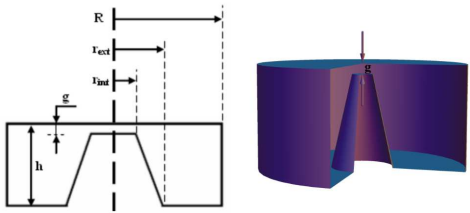}
\caption{(Color online). The klystron cavity transducer geometry (not to scale) and its relevant dimensions, and a 3D cut of a cavity shell.}
\label{cavklystron}
\end{center}
\end{figure}

The reentrant geometry for our klystron cavity is that of figure \ref
{cavklystron}. It is a truncated cone geometry, which can be machined with ease. In our case, we will use a different method for the 
critical parts of the cavity, which are near the gap region, where we need 
a surface roughness of one atomic layer. We propose to use a material with low surface roughness such as silicon, and then deposit a thin film of Niobium over it.

The field configuration for the principal mode (also known as the ``klystron 
mode", or dominant mode) is as follows. The electric field flows back and 
forth from the top of the conical post to the membrane of the cavity (which is
responsible for the cavity's top) in an area roughly equal to that of the top 
of the conical post. The magnetic field lines flow round the cavity, circling the post. 
In a first approximation \cite{HANSEN:1939p5833,HANSEN:1939p5834}, we can say that the field configuration 
near the post's top is TM$_{010}$ while the rest of the cavity is in the coaxial TEM mode. Experiments 
as well as theoretical considerations on this specific geometric arrangement 
for the klystron cavity can be found in \cite{Barroso:2004p5875, Barroso:2005p5868}.

In a lumped circuit approximation, we can roughly say that the volume between 
the cavity's top and the conical post's top (in an area approximately equal to 
that of the conical post's top) is a capacitance, and the rest of the volume 
of the cavity is mainly inductive, with a small capacitive part.

Using the Fujisawa formulas \cite{Fujisawa:1958p5806} to design a cavity under the requirements of a 10 GHz resonant frequency and a tuning coefficient
of $5$ GHz/\AA ~yields h=$2.0$ mm, R=$8.0$ mm, $r_{ext}$=$1.0$ mm, $r_{int}$=$2.7$ $\mu$m and a gap g=$1.0$ nm. The frequency scales as $f_0 \sim \sqrt{g/h}$, from which we see that the frequency selectivity
$df_0/dg \sim f_0/2g$ increases linearly with the inverse of the gap. Now, with our desired cavity with electrical parameters as described 
previously, we face the technological challenge of producing (and 
maintaining) a gap of $1$ nm. We  describe a way 
to construct our cavity and maintain the gap with the desired dimension. A gap increase would degrade the series noise, which is the main noise source for most of the band we are interested in. A further gap decrease, if possible, would enhance the backaction noise, so we can not get better sensitivities with a gap decrease. Power increase would help achieve a better sensitivity, but in principle there is an electric breakdown limit, which is discussed later in the paper.

The membrane can be devised in such a way that its first
resonant mode is at a frequency about ten times lower than the first quadrupole
mode of the sphere, so that it remains static when the sphere vibrates, and the 
gap variation is approximately equal to the sphere's radial vibration at that 
point. The upper resonance modes of the membrane have to be adjusted so that 
they do not fall into the range of frequencies of our interest, i.e., 
$3\sim10$ kHz.

The membrane's weight, together with a control system for adjustments, seems 
to be the best idea for our gap to be kept at $1$ nm. This would work as 
follows. First, we deposit the niobium film over the silicon bulk, then we close the cavity, maintaining the gap opened by
using the membraneХs weight (i.e., its natural bending by
the action of gravity) to help us. To keep the gap correctly
adjusted during operation of the cavity, we would
adjust its resonance to the klystron frequency desired ($\sim$10 GHz)
and the control system would be a piezoelectric
device, which would actuate on the gap distance by
tilting the transducer and, so, changing the force component
of the membrane weight that tries to close the gap.

LetХs now outline the main technological issues related
to the gap fabrication:

\subsection{Casimir Effect.} The Casimir effect can 
produce a force of $\sim0.03~N$ (without finite conductivity corrections), which could bend and close our gap, in an 
unstable cycle (the force bends the membrane, which reduces the gap, etc. The 
elastic properties of the gap seem to be such that this process will only 
achieve equilibrium at a bending distance bigger than the gap).

In fact this seems to be a nice device to measure the Casimir force in a new 
length scale, which has not been done yet, and can probably probe effects 
related to the granularity of matter as on this scale an atom array can not, 
in principle, be treated as continuous anymore.

With this we can take the variation 
of the energy with the gap between the plates, and we find the well-known 
expression for the force per unit area:
\begin{equation}\label{casimirforce}
\displaystyle\frac{F_{plane}(g)}{A}=-\frac{\hbar c \pi^2}{240 g^4}
\end{equation}
The negative signal means that the force is attractive.

When our physical system (in our case, the klystron cavity) has a dimension 
which is much smaller than the rest of the relevant dimensions, we are in a 
situation in which the \emph{proximity force approximation} can be summoned. It 
gives us a rough estimative of the Casimir force in our system.

If we expand the Casimir force in powers of $\hbar$, it can be shown that the first-order contribution to the force is the following. 
Suppose we take our surface and slice it in infinitesimal parallel plates of 
area $dS(g)$, separated by a distance $g$. 
Then the Casimir force will be the sum of all those contributions, i.e.
\begin{equation}
F_{PFA}=\displaystyle\int dS(g) \frac{F_{plane}(g)}{A}
\end{equation}
Where $\displaystyle\frac{F_{plane}(g)}{A}$ is given by the expression \ref{casimirforce}.  

In our particular case, it turns out that the proximity force approximation 
gives us an extremely simple way of quantifying the Casimir force in our 
cavity. As the dimensions have already been calculated, we find that the force in the gap will be $F_{PFA}\simeq3\times 10^{-2}$ N. However, by including finite conductivity corrections \cite{Lambrecht:1999p2691}, which are proportional to the size of the gap, the usual $g^{-4}$ dependence goes to $g^{-3}$. We can write $F_c= \eta_F F_{PFA}$

\begin{equation}
\eta_F=1.193 \frac{g}{\lambda_P}
\end{equation}
 
Where $\lambda_P$ is the plasma wavelength of niobium, which can be calculated from \cite{Karakozov:2004p1814}, and will give us a factor of $10^{-2}$ in the force, so $F_{c}\simeq3\times 10^{-4}$ N.

Even after manufacturing surfaces (top of the microwave cavity post and central region of the membrane) with precision of a few atomic layers and align them paralell to each other, there is still the problem to adjust the gap to one nanometer. In order to acomplish this, we have to take into acount the following forces: Casimir, membrane spring force, and gravity. The force of gravity can be used to adjust the gap, because it is a function of the tilting we give to the membrane. On the other hand, the membranes of transducers placed on the spherical antenna surface according to the truncated icosahedron configuration \cite{Johnson:1993p9003} can only have two angles of inclination with the vertical: 52.6225 degrees for the three transducers close to the ТnorthУ pole and and 10.8123 degrees for the three transducers close to the ТequatorУ. So, the force of gravity will have values around the effective weight of the membrane times the sine of these angles. The Casimir force is always negative (tries to close the gap), so, the membrane spring force needs to be positive or, in other words, the gap for which the membrane spring is relaxed should be larger then one nanometer. This Тrelaxed gapУ can be calculated by solving the equation of forces for the gap of one nanometer, taking into account the effective mass of the membrane, the resonant frequency of it, and its positioning angle mentioned above. The force of gravity must also be negative otherwise we can not go from the Тrelaxed gapУ to the one nanometer gap by changing the membrane angle from vertical (relaxed) to the positioning angle, when the transducer is placed on the sphere surface.

\begin{figure}[h]
\begin{center}
\includegraphics[width=.4\textwidth]{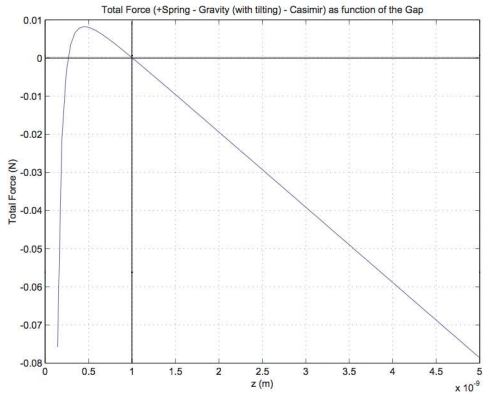}
\caption{The forces acting on the membrane and the stable point that will give the desired 1 nm gap.}
\label{stab}
\end{center}
\end{figure}

In Figure \ref{stab} it is shown the total force (membrane spring minus gravity (with tilting) and minus Casimir force) as a function of the gap. As designed, the forces should cancel for a gap of one nanometer. The Casimir force is negligible at this gap compared to the other two. However, it is the dominant one for much smaller gaps. The inclination of the linear part of the curve is given by the spring constant value. The equilibrium of forces at the gap of one nanometer is stable, because the curve has a negative derivative at the equilibrium point, which means that the positive spring force becomes the dominant when the gap departs from equilibrium into the negative direction, and the negative gravity force becomes the dominant otherwise.

Just to take a grasp at the influence of conduction electrons (that produce an attractive force in the gap), let us make a 
rough estimative of the force between two parallel plates with uniform charge 
density $\sigma$. A straightforward calculation leads us to (supposing the 
distance between the plates, $g$, is much smaller than the plates' radii, $a$)
\begin{equation}
F=\frac{\pi a^2 \sigma^2}{\epsilon_0}=\epsilon_0 \pi a^2 E^2 \label{electricforce}
\end{equation}
where we have used the approximation that the plates behave like infinite 
planes. As the electric field is variable, we will have a variable force that 
oscillates like $\sin^2(\omega_0 t)$ where $\omega_0$ is the microwave pump 
frequency. As we are working near the electric field breakdown, we can use 
the expression (Kilpatrick's criterion):
\begin{equation}
E_{break}=800 \sqrt{ f[Hz]} ~~[V/m]
\end{equation}
Inserting this in expression \ref{electricforce} and averaging over time, 
which will give us a factor of $1/2$ in the average force, we have 
$F_{ave}\sim10^{-6}$ $N$, which is neglible.

\subsection {Electric Field Breakdown.} According to the noise density expressions, we can see that by pumping more power into the cavity we can 
lower the series noise, which is the main contribution to the sensitivity 
curve of our detector. However, as the gap is very small, the electric field 
can not be arbitrarily large, thus restricting the amount of power that can 
be pumped into the cavity.

The calculation of the upper limit goes as follows. From simulations using 
the code Poisson Superfish, we see that approximately half 
of the stored energy on the electric field of the cavity is due to the field 
concentrated near the gap, where we can assume that we have a small parallel plate capacitor.
 So, from the definition of the Q-factor:
\begin{equation}
Q=\displaystyle \frac{U_{stored}}{P_{inc}}\omega_0
\end{equation}
So that, after using $U_{stored}=2CV_{max}^2$, $C=\epsilon_0~\pi r_{int}^2/g$, Kilpatrick's
criterion for the maximum electric field admissible at a certain frequency in the vacuum, $E_{max}=800\sqrt{f\mbox{[Hz]}}~\mbox{[V/m]}$, the well-known expression
$E_{max}=V_{max}/g$, and some algebra, we are lead to $P_{inc}\sim 10^{-10}~W$. 
As the gap distance and electrical properties (frequency and $df_0/dg$) increase the sensitivity of the
Schenberg detector for small gaps, the power that can be pumped is decreased, 
which decreases the sensitivity, so we have conflicting requirements.
\subsection {Surface Roughness.} We need a gap of $1~nm$ that maintains its 
dimension throughout the extension of both the cavity's top and the post top, 
which corresponds to two circular areas of radius $2.7~\mu m$.
This is why we chose silicon as the bulk material, because silicon wafers 
nowadays have roughnesses of the order of an angstrom for areas much larger 
than the one we are considering here.

Another interesting issue is the fact that silicon corrosion is well-developed 
and we can construct the post by corrosion of a silicon wafer, protecting the 
region of the post's top, as well as doing a membrane with arbitrary geometry 
and mass, by leaving one of its faces untouched by physical or chemical 
processes. We can also have a membrane with large mass, so we tackle two 
problems at the same time: the brownian noise of the membrane, which is 
reduced by its bigger mass, and the freedom to choose the geometry of the 
membrane so that we can have its oscillation modes at the frequencies we wish 
to (in our case, the first mode at $\sim1$~kHz, and the second at 
$f_1>10$~kHz).

In the literature, thin Niobium films were deposited on a quartz substrate with
the precision we need in the klystron cavity. Techniques such as electron beam deposition, which can 
achieve deposition rates for a thin film of $0.5~\AA/min$, could 
achieve our desired surface roughness.

\section{Conclusion}

In this work, we have developed an initial idea of turning the Schenberg
gravitational wave detector into a broadband detector, because its small 
bandwidth is its greatest drawback, as it is for any resonant mass gravitational 
wave detector.

Then, using the approach in \cite{Bonaldi:2006p6716}, which does not use the hypothesis of a
small bandwidth detector, we found out that our 
detector can have various small bandwidths centered at each of its quadrupole 
modes and, following a multiple sampling argument, we can 
compose the sensitivity curves of an array of spheres so that we put the 
quadrupole mode of a sphere at the frequency where the other spheres have 
poor sensitivity. This array could be mounted in such a way that, if we put one 
sphere's mode side-by-side with others' modes, we would obtain a smooth 
sensitivity curve. An example with a 6 spheres array was 
depicted in Figure \ref{sensitivityschenberg}.

A possibility of measuring the Casimir effect at the edge of the 
nanometric scale was sketched, because we need to develop a Klystron cavity 
with a $1~$nm gap wide. This poses many technological as well as theoretical 
questions, and this seems to be a rich and challenging device, both in 
its fabrication as well as pursuing a precise measurement of the Casimir 
effect in its geometry.

In the condensed matter area, we seem to have a great challenge also, trying 
to achieve a niobium film deposition over silicon with atomic layer precision. 
Several alternative techniques, such as the usage of powerful lasers and other 
alternatives to the usual sputtering process (which leaves big 
bubbles when depositing niobium, $\sim 1~\mu$m) might be tested, as well as 
thermal treatments of the bulk.

This spherical antenna array is a challenging and very deep scientific 
investigation, as well as the development of the necessary electronics, 
specially the reentrant cavity. The attractive idea is its cost, two orders 
of magnitude below the typical interferometer cost, as well as the main 
advantages of spherical detectors \cite{Forward:1971p9001, Wagoner:1977p9002, Johnson:1993p9003, Dubath:2007p14878, Cerdonio:2001p6394, Coccia:1997p9004, deWaard:2000p9005}: omnidirectionality, determination of the 
polarization of the incoming wave, as well as its direction (except for the 
sense, i.e., if we point the z-axis in the direction of the wave, we can not be sure if 
the wave is coming from the +z or -z sense), low cost, possibility for many 
smaller groups to have their own gravitational wave observatory, and thus being 
able to work with the whole hardware, software and data analysis process.

The remarkable task of detecting gravitational waves has resisted efforts 
from many groups throughout the world since the sixties. This proposal may 
lead us to a robust and highly reliable network of detectors which may help 
the ground-based interferometers ($10-100~$Hz) and space interferometers ($0.1-10~$Hz) in the band of $1-10~$kHz to open a new and certainly surprising 
window to observe the universe.

\begin{acknowledgments}
GLP would like to acknowledge Livia Conti for discussions on \cite{Bonaldi:2006p6716}, Robert L. Jaffe, Carlos Farina, Paulo A. Maia Neto and the participants of the workshop ``60 Years of the Casimir Effect" for discussions regarding the arise of the Casimir effect in the gap of the cavity, as well as David McGady. He was supported by CAPES, the Fulbright foundation, and Princeton University. This work has been supported also by FAPESP (under grant \# 2006/56041-3).
\end{acknowledgments}

\bibliography{bibliopaper}
\end{document}